\documentstyle[aps,prl,epsf,floats]{revtex} 
\draft
\begin{document}
\def\SNG{{\em Physical Review Style and Notation Guide}}
\def\LUG {{\em \LaTeX{} User's Guide \& Reference Manual}}
\def\btt#1{{\tt$\backslash$\string#1}}%
\def\REVTeX{REV\TeX}
\def\AmS{{\protect\the\textfont2
        A\kern-.1667em\lower.5ex\hbox{M}\kern-.125emS}}
\def\AmSLaTeX{\AmS-\LaTeX}
\def\BibTeX{\rm B{\sc ib}\TeX}
\title{Breakdown of Landau-Ginzburg-Wilson theory for certain quantum phase
       transitions}
\author{Thomas Vojta, D.Belitz, and R.Narayanan}
\address{Department of Physics and Materials Science Institute,
University of Oregon,
Eugene, OR 97403}
\author{T.R.Kirkpatrick}
\address{Institute for Physical Science and Technology, and Department of Physics\\
University of Maryland, College Park, MD 20742}
\date{\today}
\maketitle
\begin{abstract}
The quantum ferromagnetic transition of itinerant electrons is considered.
It is shown that the Landau-Ginzburg-Wilson theory described by
Hertz and others breaks down due to a singular coupling
between fluctuations of the conserved order parameter. This coupling
induces an effective long-range interaction between the spins of the form
$1/r^{2d-1}$. It leads to unusual scaling behavior at the quantum critical
point in $1<d\leq 3$ dimensions, which is determined exactly.
%
%
\end{abstract}
\pacs{PACS numbers: 64.60.Ak , 75.10.Jm , 75.40.Cx , 75.40.Gb} 

 
One of the most obvious examples of a quantum phase transition
is the ferromagnetic transition of itinerant
electrons at zero temperature $T$ as a function of the 
exchange coupling between the electron spins.
Hertz\cite{Hertz} derived a 
Landau-Ginzburg-Wilson (LGW) functional for this case in analogy
to Wilson's treatment of classical phase transitions, and
analyzed it by means of renormalization group methods. He found
that the critical behavior in dimensions $d=3,2$
is mean-field like,
since the dynamical critical exponent $z$ decreases the upper critical 
dimension $d_c^+$ compared to the classical case. In a quest for
nontrivial critical behavior, Hertz studied a model with a 
magnetization confined
to $d<3$ dimensions, while the
coefficients in the LGW functional are those of a $3$-$d$ Fermi
gas. For this model he concluded that $d_c^+ = 1$, and
performed a $1-\epsilon$ expansion to calculate 
critical exponents in $d<1$. Despite the artificial
nature of his model, there is a general belief that the qualitative
features of Hertz's analysis, in particular the fact that there is mean-field
like critical behavior for all $d>1$, apply to real itinerant
quantum ferromagnets as well.

In this Letter we show that this belief is mistaken, since
the LGW approach breaks down due to the
presence of soft modes in addition to the order parameter fluctuations, 
viz. spin-triplet particle-hole excitations
that are integrated
out in the derivation of the LGW functional. These soft modes
lead to singular vertices in the LGW
functional, invalidating the LGW philosophy of deriving an effective local 
field theory in terms of the order parameter only\cite{LGWfootnote}.
In Hertz's original model this
does not change the critical behavior in $d>1$, but it invalidates his
$1-\epsilon$ expansion. 
More importantly, in a more realistic model the same effect
leads to nontrivial critical
behavior for $1<d\leq 3$, which we determine exactly.

Our results for realistic quantum magnets can be summarized as follows.
The magnetization, $m$, at
$T=0$ in a magnetic field $H$ is given by the equation of state
\begin{equation}
t\,m + v\,m^d + u\,m^3 = H\quad,
\label{eq:1}
\end{equation}
where $t$ is the dimensionless distance from the critical point, and
$u$ and $v$ are finite numbers. From (\ref{eq:1})
one obtains the critical exponents $\beta$ and $\delta$, defined by
$m \sim t^{\beta}$ and $m \sim H^{1/\delta}$, respectively, at $T=0$.
For $\beta$ and $\delta$, and for
the correlation length exponent
$\nu$, the order parameter susceptibility exponent $\eta$, and the
dynamical exponent $z$, we find
\begin{equation}
\beta = \nu = 1/(d-1),\quad \eta = 3-d,\quad \delta = z = d, \quad\quad
(1<d<3),
\label{eq:2}
\end{equation}
and $\beta=\nu=1/2$, $\eta=0$, $\delta=z=3$ for $d>3$. These exponents
`lock into' mean-field values at $d=3$, but have nontrivial values for
$d<3$. In $d=3$, there are logarithmic
corrections to power-law scaling. Eq.\ (\ref{eq:1}) applies to $T=0$.
At finite temperature, we find homogeneity laws for $m$, and for the
magnetic susceptibility, $\chi_m$,
\begin{mathletters}
\label{eqs:3}
\begin{equation}
m(t,T,H) = b^{-\beta/\nu} m(tb^{1/\nu}, Tb^{\phi/\nu}, Hb^{\delta\beta/\nu})
                                                                    \quad,
\label{eq:3a}
\end{equation}
\begin{equation}
\chi_m(t,T,H) = b^{\gamma/\nu} \chi_m(tb^{1/\nu}, Tb^{\phi/\nu}, 
                           Hb^{\delta\beta/\nu})\quad,
\label{eq:3b}
\end{equation}
\end{mathletters}%
where $b$ is an arbitrary scale factor. The exponent $\gamma$, defined by
$\chi_m \sim t^{-\gamma}$ at $T=H=0$ and the crossover exponent $\phi$ 
that describes the crossover
from the quantum to the classical Heisenberg fixed point (FP) are given by
\begin{equation}
\gamma = \beta (\delta - 1) = 1 \quad,\quad \phi = \nu \quad,
\label{eq:4}
\end{equation}
for all $d>1$. Notice that the temperature dependence of the
magnetization is {\em not}
given by the dynamical exponent. However, $z$ controls the temperature 
dependence of the specific heat coefficient, $\gamma_V = c_V/T$, which
has a scale dimension of zero for all $d$, and logarithmic corrections
to scaling for all $d<3$\cite{Wegner},
\begin{equation}
\gamma_V(t,T,H) = \Theta(3-d)\,\ln b +
           \gamma_V(tb^{1/\nu}, Tb^z, Hb^{\delta\beta/\nu})\quad .
\label{eq:5}
\end{equation}
Eqs. (\ref{eq:1}) - (\ref{eq:5}) represent the exact critical
behavior of itinerant quantum Heisenberg ferromagnets for all
$d>1$ with the exception of $d=3$, where additional logarithmic
corrections to scaling appear.
We are able to obtain the critical behavior exactly, yet it is not
mean-field like. The exactness is due to the fact that we work above
the upper critical dimension $d_c^+ = 1$. The nontrivial exponents are
due to a singular coupling between the critical modes which
leads, e.g., to the unusual term $\sim v$ in (\ref{eq:1}). Experimentally,
we predict that for $3$-$d$ magnets with a very low $T_c$ there is a 
crossover from essentially mean-field quantum behavior to classical Heisenberg
behavior. In $d=2$, where there is no classical transition, we predict that
with decreasing $T$, long-range order will develop, and the quantum phase
transition at $T=0$ will display the nontrivial critical behavior shown
above. 

We now sketch the derivation of these results.
A more complete account of the technical details will be
given elsewhere \cite{ustbp}.
We consider a $d$-dimensional continuum model of interacting
electrons, and pay particular attention to the particle-hole spin-triplet
contribution \cite{AGD} to the
interaction term in the action, $S^t_{\rm int}$, whose (repulsive) coupling
constant we denote by $J$. Writing only the latter
explicitly, and denoting the spin density by ${\bf n}_s$, the action reads,
\begin{equation}
S = S_0 + S_{\rm int}^t
  = S_0 + (J/2) \int dx\ {\bf n}_s(x)\cdot{\bf n}_s(x)\quad,
\label{eq:6}
\end{equation}
where $S_0$ contains all contributions
to the action other than $S_{\rm int}^t$. In particular, it contains the
particle-hole spin-singlet and particle-particle interactions, 
which will be important for what
follows. $\int dx = \int d{\bf x} \int_0^{1/T} d\tau$, and we use
a 4-vector notation $x = ({\bf x}, \tau)$, with ${\bf x}$ a vector
in real space, and $\tau$ imaginary time. Following Hertz, we perform a
Hubbard-Stratonovich decoupling of $S_{\rm int}^t$ by introducing a
classical vector field ${\bf M}(x)$ with components $M^i$ that couples to 
${\bf n}_s(x)$ and whose
average is proportional to the magnetization, and we
integrate out all fermionic degrees of freedom. 
We obtain the partition function $Z$ in the form
\begin{mathletters}
\label{eqs:7}
\begin{equation}
Z = e^{-F_0/T} \int D[{\bf M}]\,\exp\bigl[-\Phi[{\bf M}]\bigr]\quad,
\label{eq:7a}
\end{equation}
where $F_0$ is the noncritical part of the free energy.
The Landau-Ginzburg-Wilson (LGW) functional $\Phi$ reads
\begin{eqnarray}
\Phi[{\bf M}] = {1\over 2} \int dx\,dy\ {1\over J}\delta(x-y) 
{\bf M}(x)\cdot {\bf M}(y)
+ \sum_{n=2}^{\infty}a_n \int dx_1\,\ldots\,dx_n\ \chi^{(n)}_{i_1\ldots i_n}
  (x_1,\ldots,x_n) 
M^{i_1}(x_1)\,\ldots\,M^{i_n}(x_n)\ ,
\label{eq:7b}
\end{eqnarray}
\end{mathletters}%
where $a_n = (-1)^{n+1}/n!$. The coefficients
$\chi^{(n)}$ in (\ref{eq:7b}) are connected n-point spin density
correlation functions of a
reference system with action 
$S_0$\cite{Hertz}. 
The particle-hole spin-triplet interaction $J$ is missing in the bare
reference system, but a nonzero $J$ is generated perturbatively by the
particle-particle interaction contained in $S_0$.
The reference system then has all of the characteristics of 
the full action $S$,
except that it must not undergo a phase transition lest the separation of
modes that is implicit in our singling out $S_{\rm int}^t$ for the
decoupling procedure breaks down.

$\chi^{(2)}$ is the spin susceptibility of the
reference system. Performing a Fourier
transform from $x=({\bf x},\tau)$ to $q=({\bf q},\Omega)$ with wavevector
${\bf q}$ and Matsubara frequency $\Omega$, we have for small ${\bf q}$ and
$\Omega$\cite{limitsfootnote},
\begin{mathletters}
\label{eqs:8}
\begin{equation}
\chi^{(2)}({\bf q},\Omega)
           = \chi_0({\bf q})[1 - \vert\Omega\vert/\vert{\bf q}\vert]\quad,
\label{eq:8a}
\end{equation}
where ${\bf q}$ and $\Omega$ are being measured in suitable units, and
$\chi_0({\bf q})$ is the static spin susceptibility of the reference
system. We now use the fact that in a Fermi liquid at $T=0$,
$\chi_0$ is a nonanalytic function of ${\bf q}$ of the form
\begin{equation}
\chi_0({\bf q}\rightarrow 0) \sim {\rm const} - \vert{\bf q}\vert^{d-1}
                                                - {\bf q}^2 \quad.
\label{eq:8b}
\end{equation}
\end{mathletters}%
Here we have omitted all prefactors, since they are irrelevant for our
purposes. This holds for $1<d<3$; in $d=3$ the nonanalyticity is of the form 
${\bf q}^2 \ln\vert{\bf q}\vert$\cite{qd-1footnote}. Using (\ref{eqs:8}),
and with $\int_q = \sum_{\bf q} T\sum_{i\Omega}$,
the Gaussian part of $\Phi$ can be written,
\begin{equation}
\Phi^{(2)}[{\bf M}] = \int_q {\bf M}(q)\bigl[t_0
                       + c_n\vert{\bf q}\vert^{d-1} + c_a{\bf q}^2
                + c_d\vert\Omega\vert/\vert{\bf q}\vert\bigr]\,
                                                 {\bf M}(-q)\quad.
\label{eq:9}
\end{equation}
Here $t_0 = 1 - \Gamma_t\chi^{(2)}({\bf q}\rightarrow 0,\omega_n = 0)$
is the bare distance
from the critical point, and $c_n$, $c_a$ and $c_d$ are constants.

For the same physical reasons for which the nonanalyticity occurs in
(\ref{eq:8b}), the coefficients $\chi^{(n)}$ in (\ref{eq:7b})
are in general not finite
at zero frequencies and wavenumbers. Let us focus in $\chi^{(4)}$, which will
be the most interesting one for our purposes. Again, standard 
perturbation theory shows
that it is given schematically by\cite{ustbp}
\begin{equation}
\chi^{(4)} \sim {\rm const} + v\int_k \bigl[\vert{\bf k}\vert
                            + \vert\omega_n\vert\bigr]^{-4}
           \sim u +  v\vert{\bf p}\vert^{d-3}\ .
\label{eq:10}
\end{equation}
Here we have cut off the singularity by means of a wave\-number
$\vert{\bf p}\vert$, and $u$ and $v$ are finite numbers.
More generally, the coefficient of $\vert{\bf M}\vert^{n}$ in $\Phi$
for $\vert{\bf p}\vert\rightarrow 0$
behaves like $\chi^{(n)} = v^{(n)} \vert{\bf p}\vert^{d+1-n}$.
This implies that $\Phi$ contains a nonanalyticity
which in our expansion takes the form of a power series in
$\vert{\bf M}\vert^2/\vert{\bf p}\vert^2$.

The functional $\Phi$ can be analyzed by using standard
techniques \cite{Ma}. We look for a FP where $c_d$ and either $c_n$
(for $1<d<3$), or $c_a$ (for $d>3$) are not renormalized. This fixes the
critical exponents $\eta$ and $z$. Choosing the scale dimension of a length
$L$ to be $[L] = -1$, standard power counting\cite{Ma} then yields the
scale dimension of $v^{(n)}$ to be $[v^{(n)}] = -(n-2)(d-1)/2$. All
non-Gaussian terms are thus irrelevant for $d>1$, and they all become
marginal in $d=1$ and relevant for $d<1$.
Several features of the critical
behavior follow immediately. The
critical exponents $\eta$ and $z$ are fixed by the choice of our FP, and
$\nu$ and $\gamma$ as given in (\ref{eq:2}) and (\ref{eq:4}) are
obtained by considering the
${\bf q}$-dependence of the Gaussian vertex (\ref{eq:9}).
We determine the equation of state by taking
the term of order $\vert{\bf M}\vert^4$ in $\Phi$ into account.
$\chi^{(4)}$ is dangerously irrelevant with respect to the 
magnetization. We have shown \cite{ustbp} that for scaling purposes
the cutoff $\vert{\bf p}\vert$ in (\ref{eq:10}) can be replaced by
$m$. From this and (9) we obtain 
the effective equation of state as given in (\ref{eq:1}).

These results completely specify the critical behavior at $T=0$. Their
most interesting aspect is the nontrivial exponent values found for
$1<d<3$, which can nevertheless be determined exactly. The reason for this is
the $\vert{\bf q}\vert^{d-1}$-term in the Gaussian action (\ref{eq:9}).
It reflects the fact that in an interacting electron system,
static correlations between spins do not fall off exponentially with
distance, but only algebraically like $r^{-(2d-1)}$.
This slow decay leads to a
long-range interaction in the effective action
which falls off like $1/r^{2d-1}$, see (\ref{eq:9}). The critical
behavior of {\em classical} Heisenberg magnets with such a long-range
interaction has been studied before \cite{FisherMaNickel}.

We now turn to the
$T$-dependence of the specific heat, $c_V$.
We expand the free energy functional (\ref{eq:7b})
about the expectation value, $m$, of ${\bf M}$ to second
order, and then perform the Gaussian integral to obtain the partition
function. The free energy is obtained as the sum of
a mean-field contribution given by $\Phi[m]$, and a fluctuation contribution
given by the Gaussian integral. The latter yields the leading nonanalytic
term in the free energy.
We find \cite{ustbp} that effectively $H$ and $T$ have the
same scale dimension, viz. $d (=z)$, and that at $t=0$ there is
a logarithmic $T$-dependence of $\gamma_V$ for {\em all} $1<d<3$.
If we put the $t$-dependence back in,
we obtain that the scale dependence of $\gamma_V$ is given by
(\ref{eq:5}).

For the magnetization the leading $T$-dependence
is given by the mean-field contribution to the free energy.
We calculate the temperature corrections to the equation of state 
(\ref{eq:1}) and find that for $m>>T$ (in suitable units) $m^d$ in
(\ref{eq:1}) will be replaced by
$m^d[1+{\rm const}\times T/m + \ldots]$,
while for $m<<T$, $t$ is replaced by $t + T^{1/\nu}$. The effective scale
dimension of $T$ in $m$ is therefore $1$ ({\em not} $z$), and we obtain
for $m$ and $\chi_m$ the homogeneity laws given by (\ref{eqs:3}).
Thus, the relevant operator $T$
in (\ref{eqs:3}) reflects the crossover from the quantum 
to the classical FP rather than dynamical scaling.
Accordingly, we have written the $T$-dependence
in (\ref{eqs:3}) in terms of a crossover exponent $\phi$
which is given by (\ref{eq:4}).

Next we briefly discuss Hertz's original model, which
differs from the one discussed above in two ways. First,
the reference ensemble consists of noninteracting electrons.
Second, the coefficients $\chi^{(n)}$
are taken to be the correlation
functions of a $3$-$d$ fermion system.
$\chi^{(2)}$ is then simply the Lindhard function, so
(\ref{eqs:8}) gets replaced by,
\begin{equation}
\chi^{(2)}({\bf q},\Omega) = 1 - {\bf q}^2 
             - \vert\Omega\vert/\vert{\bf q}\vert + \ldots\quad.
\label{eq:11}
\end{equation}
Due to the missing interaction in the reference ensenble,
$\chi^{(2)}({\bf q},0)$ is now analytic at $\vert{\bf q}\vert=0$.
The resulting quadratic term in (\ref{eq:7b})
allows for a Gaussian FP with mean-field static exponents
and a dynamical exponent $z=3$\cite{Hertz}. Whether this FP is
stable depends on the higher $\chi^{(n)}$.
Hertz considered only the limit ${\bf q} = \Omega = 0$,
where all of these terms are finite numbers and irrelevant
for $d>1$. The quartic term is marginal in $d=1$ and
relevant for $d<1$\cite{Hertz}.
 
The striking difference between the finite coefficients in Hertz's model
and the diverging ones in the realistic model above is due
to the latter containing interactions in the reference ensemble. The
interactions lead to frequency mixing, and hence to soft particle-hole
excitations contributing to the $\chi^{(n)}$ even in the limit of
zero external frequency. A similar effect is achieved for noninteracting
electrons by considering correlation functions at nonvanishing external
frequency. Therefore we include
the higher order terms in an expansion of the $\chi^{(n)}$
in powers of $\Omega$ and analyze the arising LGW functional by the
same power counting arguments as above. The details of this calculation
will be presented elsewhere [4]. We find
that all non-Gaussian terms are still irrelevant for
$d \geq 2$ and the critical behavior is mean-field like.
In $d=1$, however, the $\chi^{(n)}$ change their
functional form so that an infinite number of operators is relevant ({\em not} 
marginal) with respect to the Gaussian FP in $d = 1$ and below.
Therefore the upper critical dimension is {\em not}
one, but rather the $1$-$d$ sytem is {\em below} its upper critical
dimension, and will show critical behavior that is substantially different
from mean-field behavior. 
 
We conclude with a few remarks. 
First, the vertices in the LGW functional discussed here
are singular only if the order parameter is conserved, and only
at zero temperature, which means
that the phenomenon is confined to the quantum magnetic transition.
Second, our conclusion, although derived for the
special example of itinerant quantum ferromagnetism, is rather general:
We expect the LGW formalism to break down whenever there are soft modes
other than the critical order parameter fluctuations that couple to the
order parameter. The general rule is that {\em all} of the soft modes should be
retained on equal footing in the effective theory. If any of them are
integrated out, the resulting penalty are ill-behaved coefficients in
the LGW functional. This has been shown recently for {\em disordered}
electrons\cite{fm}. The present results indicate that
the underlying principle is very general. Indeed, it also applies
to classical phase transitions with additional soft modes. 
However, there are many modes that are
soft at $T=0$ but acquire a mass at finite temperature, making quantum
phase transitions more likely candidates.
Finally, we mention that Sachdev\cite{Sachdev} has noted that something must be
wrong with Hertz's theory in $d<1$, since it violates an exact exponent
equality for quantum phase transitions with conserved order parameters.
He suspected Hertz's omission of the cubic term in the LGW functional to
be at fault. Our analysis provides instead the 
explanation given 
above, namely the presence of {\em infinitely} many relevant operators due 
to the soft particle-hole excitations.
 
This work was supported by the NSF under grant Nos. DMR-92-17496 and
DMR-95-10185, by the DAAD, by the DFG
under grant number Vo 659/1-1, and by the NATO under grant No. CRG-941250.
DB would like to thank Bernhard Kramer at the University of Hamburg
for hospitality.

\vfill\eject

\begin{references}
\bibitem{LGWfootnote} We use the term `LGW theory' in the narrow sense, in
 which it is usually used in the literature, of
 an effective field theory in terms of the order parameter field only.
\end{references}
\end{document}